# An All-Electric Single-Molecule Hybridisation Detector for short DNA Fragments

A.Y.Y. Loh,[1] C.H. Burgess,[2] D.A. Tanase,[1] G. Ferrari,[3] M.A. McLachlan,[2] A.E.G. Cass,[1] T. Albrecht*[1,4]

[1]Imperial College London, Department of Chemistry, Exhibition Road, London SW7 2AZ, UK

[2]Imperial College London, Department of Materials and Centre for Plastic Electronics, London SW7 2AZ, United Kingdom

[3] Politecnico di Milano, Dipartimento di Elettronica, Informazione e Bioingegneria, P.za Leonardo da Vinci 32, Milano, Italy

[4]University of Birmingham, School of Chemistry, Edgbaston Campus, Birmingham B15 2TT, UK

* t.albrecht@bham.ac.uk

**Abstract**

In combining DNA nanotechnology and high-bandwidth single-molecule detection in nanopipettes, we demonstrate an all-electric, label-free hybridisation sensor for short DNA sequences (< 100 nt). Such short fragments are known to occur as circulating cell-free DNA in various bodily fluids, such as blood plasma and saliva, and have been identified as disease markers for cancer and infectious diseases. To this end, we use as a model system a 88-mer target from the RV1910c gene in *Mycobacterium tuberculosis* that is associated with antibiotic (isoniazid) resistance in TB. Upon binding to short probes attached to long carrier DNA, we show that resistive pulse sensing in nanopipettes is capable of identifying rather subtle structural differences, such as the hybridisation state of the probes, in a statistically robust manner. With significant potential towards multiplexing and high-throughput analysis, our study points towards a new, single-molecule DNA assay technology that is fast, easy to use and compatible with point of care environments.

Nanopore devices are a new class of stochastic single-molecule sensors. As nanoscale analogues of the well-known Coulter counter, which is routinely used for cell counting in hospital environments, they have been developed towards fast and label-free DNA sequencing.[1] This feat has now largely been achieved with (modified) biological pores, such as $\alpha$-hemolysin.[2] However, resistive pulse sensing with solid-state nanopores and nanopipettes offers a range of other potential applications. These nanodevices are relatively easy to fabricate (especially nanopipettes[3,4]) and there is usually considerable flexibility in their design, with regards to the pore dimensions (diameter, channel length, shape). This means that they can more readily be adapted to larger or structurally more complex analytes, including double-stranded (ds) DNA, peptide nucleic acid (PNA)/DNA or protein/DNA complexes, and potentially be used as an 'all electric' sensor concept in gene profiling or fingerprinting, for disease diagnostics and monitoring.[5,6,7,8,9,10]

The general operating principle is rather simple, as illustrated in fig. 1 A) and explained in detail elsewhere.[8] Briefly, in a nanopore device the pore channel is typically the largest source of electric resistance in the cell. When an ion current is driven through the system *via* an applied voltage $V_{bias}$, any changes in the pore resistance thus result in a measureable change in the ion current *I* through the system. This occurs, for example, when DNA, charged particles or proteins pass through the channel.[11,12,13,14] The ion current modulation can be low (~100 pA) and short-lived (< 1 ms), depending on the analyte, the pore design and the experimental conditions. In a simple case, for example involving a cylindrical pore channel, the corresponding *I*(*t*) modulation (translocation 'event' with a duration $\tau_e$) is approximately rectangular in shape, but sub-structure is usually found for more complex analytes. For example, a protein bound to DNA typically produces an individual spike ('sub-event' with a duration $\tau_{se}$) in the *I*(*t*) trace that is superimposed on the actual DNA translocation event.[15] Hence, the number and relative positions of the sub-events can thus provide information on the number of bound proteins, potentially the thermodynamics of the binding equilibrium and the location of the proteins along the strand (if the translocation speed is known). Since the sub-event duration is normally small compared to the event duration, $\tau_{se} \ll \tau_e$, resolving the sub-events electrically can be challenging, and requires the detection of rather low currents at high bandwidth. However, recent developments in instrument design now routinely allow for time resolutions well below 10 μs with nanopipettes and even lower with nanopore chips.[16,17,18,19]

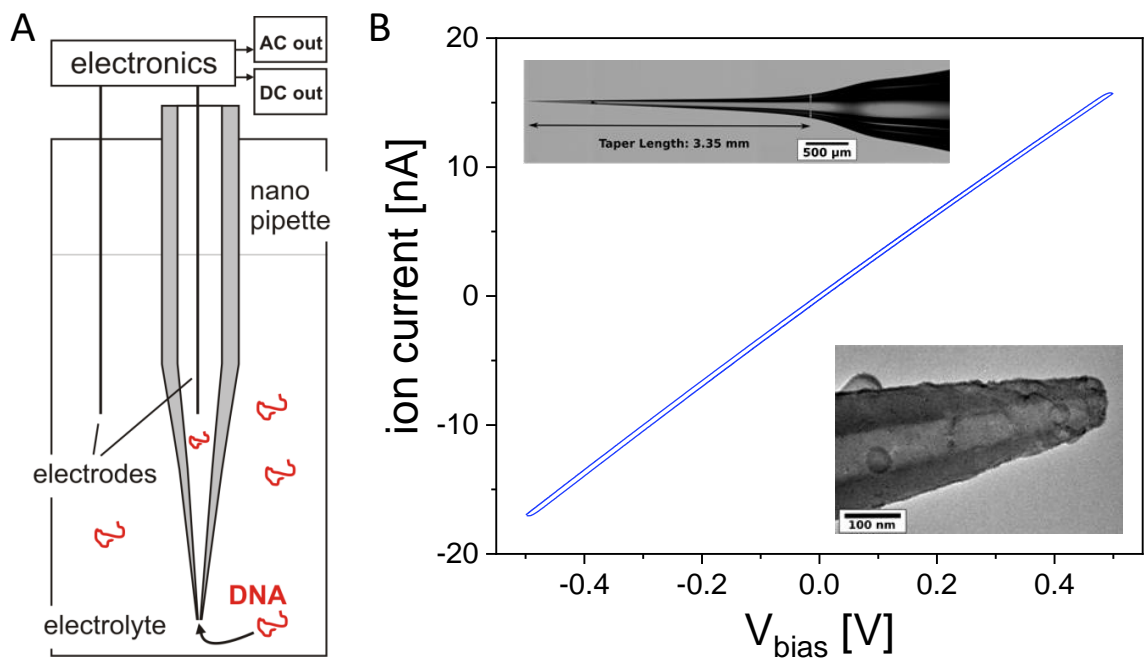

**Fig. 1**: A) Illustration of the experimental setup, cross-sectional view (not to scale). The quartz nanopipette is immersed in a liquid-filled cell, typically containing a highly concentrated chloride solution as the electrolyte. In our experiments, DNA was translocated from the outside to the inside of the pipette, as indicated. The custom-built detection electronics split the pore current into a slow ('DC') and a fast ('AC') channel, where the former contains the average pore current and the latter the translocation events. B) $I/V_{bias}$ curve for a typical quartz nanopipette in 1 M KCl + 10 mM TE buffer. The conductance is $G$ = 33.1 nS, as determined from the average of the slopes from the forward and reverse scan between +/- 0.1 V. The rectification ratio RR is 0.98 for these voltages. Top inset: Optical micrograph showing the overall shape of the same pipette. Bottom inset: TEM image of the pipette's tip. The blob-like features are built up with imaging time and are most likely due to carbon contamination. The long taper length and conical shape especially towards the pipette tip are apparent.

Meller *et al.* pioneered this concept with chip-based nanopores (diameter ~ 5 nm, $Si_3N_4$ membrane) and two different types of PNA (specifically, bis-PNA and γ-PNA), to probe short base sequences in long pieces of dsDNA and ultimately genes.[6,7] PNA binds to dsDNA in a sequence-specific manner and with very high affinity, resulting in a local change in structure (bulging). The latter in turn produces sub-structure in the translocation data, which can be related to the presence (or absence) of a particular gene sequence. Meller *et al.* exploited this capability to differentiate sub-types of the HIV pol-1 gene, for pathogen classification.

A conceptually different approach is to employ artificial, engineered structures as 'carriers' that have some function - such as protein binding capability or a recognition element - engineered into them.[20,21] For example, Bell and Keyser used nanopipettes and a carrier design based on DNA self-assembly, to include, firstly a sequence of structural features (dumbbells) as a 'barcode' identifying the DNA, and secondly a site for antibody binding.[22] Different carriers may thus be identified in mixtures and several different proteins can be assayed at the same time (multiplexing), as the authors demonstrate with biotin, bromodeoxyuridine, puromycin and digoxygenin modifications as antigens and their respective antibodies. Notably, similar engineered structures have also been used to characterise the translocation process itself, such as the translocation velocity and dynamics.[23,24], or towards the detection of single nucleotide polymorphisms.[25]

A similar idea, albeit based on aptamers, was pursued by Edel, Ivanov et al.[26] Specifically, the single-stranded (ss) ends of λ-DNA were modified with probes containing two parts: one complementary to the ss ends and a second one made of aptamer sequences optimised for thrombin and actylcholinesterase binding. This yielded DNA constructs with protein binding sites on either end, which were again analysed by translocation through nanopipettes (from the inside to the outside of the pipette in this case). The DNA carriers in earlier studies had to be fabricated by reconstituting the dsDNA from a long ss template and a large number of short, complementary strands, which is rather cumbersome and comparatively expensive. The approach of Edel and co-workers is somewhat simpler, as it starts with intact λ-DNA, albeit at the cost of reduced design flexibility and probe density. The authors also demonstrate translocation experiments in diluted human serum, which is a step towards the application of nanopore sensing in complex, perhaps more realistic media. That said, when coupled to suitable workflows, operation in such environments might not always be required.

Beamish, Tabard-Cossa and Godin combine some of the above concepts in their recent work.[27] Using sub-5 nm pores in chip-based nanopore devices (SiN membrane, thickness ~ 10 nm), prepared not by electron or ion beam drilling but by dielectric breakdown,[28] they employed DNA engineering to synthesize 255 basepair (bp) dsDNA scaffolds with ds overhangs as short as 15 bp. These overhangs could reliably be detected and resolved by ion current sensing in a label-free manner. Moreover, the authors also prepared scaffolds with ssDNA overhangs, which could bind an aptamer-based probe in the presence of ATP. The bound probe was then detected by nanopore sensing, as an indirect way of detecting ATP.

An interesting alternative approach for detecting hybridisation of short DNA (and potentially other) targets is the use of modified nanoparticles. In particular, particles with magnetic cores can first be released into the sample medium, where they bind their targets, and then re-captured and pre-concentrated using magnetic fields. Binding to the target then either changes the surface properties of the particles (e.g., zeta potential), and hence their translocation characteristics (speed),[29,30] or produces altogether new structures (such as particle dimers),[31] which are then detected by resistive pulse sensing. While these approaches do not probe individual binding or hybridisation sites, there appears to be some potential for multiplexed detection, for example by employing particles of different sizes. Apart from simple target capture, such studies have also included site-specific detection of methylation sites.[32]

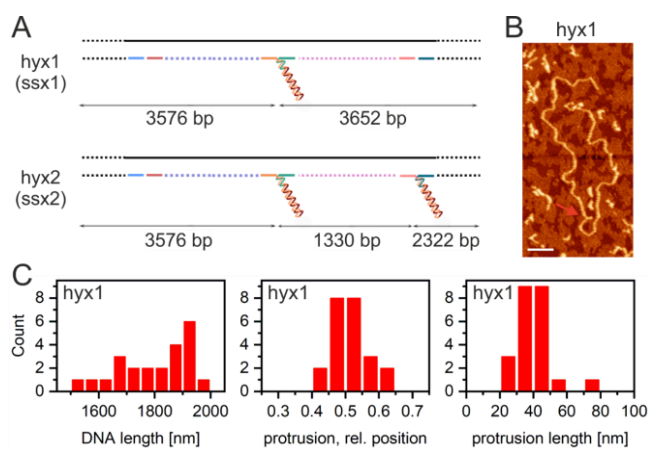

**Fig. 2**: Design and initial characterisation of the DNA structures under study. A) Basic design of the samples with one and two overhangs ('protrusions' with suffices '1' and '2', respectively) and the positions indicated. 'hyx': hybridised overhang; 'ssx': single-stranded overhang. The ss part of the overhang and its complementary target strand are ~88 nt long. B) Typical AFM image of hyx1 in air (tapping mode) after drop-casting on mica (scale bar: 100 nm). The overhang is indicated with the red arrow. The short DNA fragments are impurities from the assembly process, as shown in the gel chromatography data (SI, section 1) and nanopore translocation data below. C) Histograms of the DNA carrier length, the relative position of the protrusion and its length (based on 23 DNA structures in total).

In our present work, we build on these advances and have developed a high-throughput sensing concept with new capabilities and applications, namely with focus on the label-free detection and quantification of short (~100 nucleotide (nt)) ssDNA fragments. Such short ssDNA segments are found in blood, urine and other bodily fluids as circulating cell-free DNA (cfDNA), where they have been implicated in disease diagnostics and monitoring, for example in the context of urinary tract infections.[33] In cancer the ratio of short (< 150 bp) vs. long DNA in plasma is increased, most likely due to enhanced rates of cell apoptosis and necrosis.[34,35] Equally, such short DNA fragments may serve as diagnostic markers for infections, such as with *Mycobacterium tuberculosis* (TB)[36] and other diseases. In the present proof-of-concept study, we take TB as a model system and show how resistive pulse sensing in combination with suitable carrier design may be used to detect and potentially quantify TB-relevant, short DNA sequences in solution. Specifically, we designed 7.2 kbp long dsDNA structures with either one or two protrusions ('overhangs') in specific locations along the carrier strand (see Methods section and SI for details). These protrusions comprised of a short (~12 bp) ds section close to the carrier backbone and an 88 nt long ss section, which could be hybridized with its complementary sequence (the target). In translocation experiments with quartz nanopipettes, we then detected and differentiated ss and hybridised protrusions and hence determined the hybridisation state of the overhang in a rapid, 'all-electric' and label-free manner. While the present study focuses on carrier DNA with only a small number of protrusions, it is conceptually important to note that their number can be increased significantly, illustrating the method's potential for multiplexing. For example, based on theoretical considerations, it has been estimated for very similar DNA that features spaced as little as 90 nm along the carrier strand should be distinguishable by resistive pulse sensing in a nanopipette.[8] This is in good agreement with experimental results for biotin-streptavidin-labelled DNA, where the minimum resolvable distance was found to be approximately 200 bp (68 nm, assuming 0.34 nm/bp).[37] While the practical upper limit of the overhang density is currently unknown, it is conceivable that 7.2 kbp DNA could carry 20 or more protrusions, depending on details of the carrier design (e.g., sequence composition and protrusion length). The probe regions could in principle carry identical sequences, which would allow for repeated detection of the same target during carrier translocation and improved sensitivity. Alternatively, the probe regions could be selected to detect a number of molecular targets at the same time, for example to construct suitable marker panels. Finally, different carriers in mixed samples could be identified via their own translocation characteristics, for example if they have different lengths or identified by a fingerprint region.[22]

In the present case, we chose the sequence of the probe regions to be identical for all samples, also to allow for a comparison between different overhang locations. It was taken from the RV1910c gene in TB, a gene regulatory region of the *KatG* protein.[38] *KatG* is a catalase peroxidase, which is responsible for activating Isoniazid (INH), one of the most effective and specific anti-tuberculosis drug since its introduction in 1952.[39,40] Deregulation of the *KatG* gene thus triggers INH resistance in the bacteria and renders the drug useless. Furthermore, INH resistance is often the first step towards multi-drug resistance,[41] so robust and fast identification of antibiotic resistance could inform already the early stages of therapy. Based on our results, it appears that resistive pulse sensing with nanopores and nanopipettes could help address this need, in particular with further improvements in carrier and sensor design, and when coupled with a suitable workflow for sample extraction and amplification.

**Results and Discussion**

The quartz pipettes used in this study are produced with a mechanical puller and the exact geometry of the channel and the pore size at the pipette can vary to some degree. Based on conductance measurements, optical and transmission electron microscopy (TEM) imaging, we found however that the device-to-device variation for the pulling parameters used was relatively small (see Methods section). We typically obtained pipettes with (inner) pore diameters at the tip between 20 and 30 nm and good agreement between the different characterisation methods. As an example, we show the current-voltage *(I/V)* characteristics (forward and reverse voltage sweep), optical microscopy image and TEM image recorded for the same pipette in fig. 1 B). From the average slope of the two sweeps between +/- 0.1 V, a conductance of 33.1 nS was obtained, which in conjunction with equation S1 in the Supporting Information, was used to estimate a pore diameter of 29 nm. This compares well with the pore diameter determined by TEM, which yielded 31 nm for the same pipette. The TEM and optical images also reveal that the channel geometry is approximately conical over long distances with an opening angle of about 15°. The small offset between the forward and the reverse voltage scan is due to capacitive charging of the system, as discussed in detail elsewhere.[42,43]

The DNA structures under study here are illustrated in fig. 2 A) and comprise of two pairs of samples, as mentioned above. Namely, these are two structures with a single overhang ('ssx1' and 'hyx1') and two structures with two overhangs ('ssx2' and 'hyx2'). 'ssx' refers to DNA carriers with single-stranded overhangs, 'hyx' to those where the overhang(s) have been hybridised with an 88 nt complementary strand (as the model disease marker). Based on equilibrium binding considerations and taking into account the concentration conditions during the assembly, we found that the affinity of the complementary strand was high enough to ensure near quantitative binding for the 'hyx' samples (see SI, section 1).

In order to confirm that the assembly had been successful, atomic force microscopy (AFM) characterisation (tapping mode, in air) was performed in selected cases, as shown for hyx1 in fig. 2 B). A small amount of shorter adsorbed fragments is also seen, which were still present in the sample solution. In this context, we felt that further purification unnecessary, in light of the fact that in mixtures the longer DNA carrier can readily be distinguished from shorter fragments, both in AFM and nanopore sensing (see below). This is clearly a strength of the nanopore sensor concept presented here, which can ultimately simplify workflows in real-life applications.

In terms of the structural analysis of the hyx1 species on the surface, we focused on a region-of-interest of between 0.3 and 0.7 of the total DNA length and excluded features that had markedly different contrast than the DNA carrier itself (pointing to coiling, knotting or random co-adsorption of shorter DNA fragments). With regards to the DNA length, we found a rather broad distribution with an average of 1.85 ± 0.13 μm, cf. fig. 2 C) (left panel), which is somewhat shorter than the expected value of 2.46 μm (7228 bp as per design, 0.34 nm/bp). This has been observed by others before and is most likely due to the DNA on the surface not being fully stretched.[23] In support of this hypothesis, we found good agreement with the intended design for the relative position of the overhang (0.51 ± 0.05 vs. expected 0.51) and its length (39 ± 7 nm vs. expected: 34 nm), as shown in the middle and right panel of fig. 2 C). Hence, partial decomposition is unlikely the reason for the shorter observed carrier length, unless it affects the carrier symmetrically on both sides.

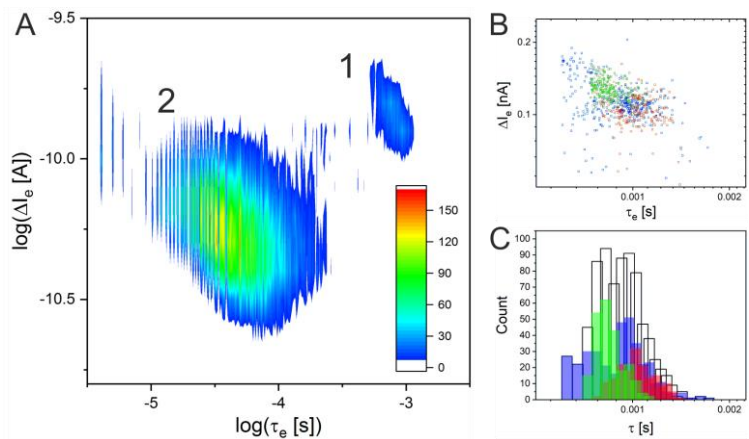

**Fig. 3**: Translocation data from three different nanopipettes for hyx1 ($V_{bias}$ = 0.7 V, 4 M LiCl + 10 mM TE electrolyte). A) Scatter density plot of $\log(\Delta I_e)$ vs. $\log(\tau_e)$. Two event clusters 1 and 2 emerge, where cluster 1 contains events from hyx1 (1166 out of a total of 59964 events). These are of interest in the present context. Cluster 2 contains shorter DNA fragments that are still present in the sample, as discussed in the context of the AFM results above. B) Scatter plot for cluster 1 only with the data points from the different pipettes colour-coded. Some small, but systematic differences arise for the cluster centres. C) One-dimensional $\tau_e$ histogram, showing the combined dataset for cluster 1 (black, solid line) as well as the individual data for each pipette (same color-coding as in B). Comparison with translocation data from ssx1, i.e. with the unhybridised overhang, reveals that the data are identical within experimental error, cf. fig. S6. This suggests that the hybridisation state of the overhang does not significantly affect the translocation characteristics of the carrier DNA.

Figure 3 shows the results of translocation experiments performed with the same 'hyx1' sample in three different nanopipettes, each with (internal) pore diameters between 20 and 30 nm ($V_{bias}$ = 0.7 V; 4 M LiCl + 10 mM TE electrolyte). In panel A), all events (59964) are combined in one (logarithmic) scatter density plot, $\log_{10}(\Delta I_e)$ vs. $\log_{10}(\tau_e)$. These events include electric noise and the translocation of short DNA fragments at short $\tau_e$, as well as the translocation of the DNA carriers at longer $\tau_e$. Two clusters, labelled '1' and '2', clearly emerge where cluster 1 contains the translocation events from the longer DNA carriers (1166 events), in line with the translocation characteristics reported for similar DNA under comparable conditions.[16] Panels B) and C) show a blow-up of cluster 1 and one-dimensional histograms, respectively, with the data color-coded according to the pipette used. The histogram in white/black solid line combines all three datasets. It is well represented well by a log-normal fit with a mean translocation time of ‹$\tau_e$› = 0.94 ± 0.01 ms. However, as shown by the color-coded individual datasets, there are small, but systematic differences between the individual nanopipettes used. This is not surprising, since the channel dimensions are known to affect the translocation time and the associated current modulation.[44,8] Specifically, for larger pore diameters $d_p$, $\tau_e$ and $\Delta I_e$ (relative to mean pore current) decrease, for smaller $d_p$, the opposite effect is observed. So, while every effort was made to use very similar nanopipettes in the experiments, in terms of their conductance $G$, the actual pore dimensions, and thus the translocation characteristics of an analyte, will not be exactly the same. The weighted average of the translocation times for all three pipettes is ‹$\tau_e$› = 1.0 ± 0.2 ms (weighted standard error) and hence the same as the previous value of the mean translocation time, within experimental error. The small relative shifts between the individual translocation time distributions, however, led to some broadening of the overall (combined) translocation time distribution, which would in turn affect the determination of related parameters, such as the effective diffusion coefficient of the DNA segment in the pore.[45] However, this aspect is not in focus of the present study and we now turn to the discussion of the event sub-structure, related to the presence of the different overhangs.

Three example events of each case – ssx1, hyx1, ssx2 and hyx2 – are shown in fig. 4 A) ($V_{bias}$ = 0.7 V), along with graphical illustration of some parameters used for further analysis (cf. Methods section). ssx samples are colored in green, hyx samples in red throughout this figure. Analogous data recorded at $V_{bias}$ = 0.5 V using different pipettes are shown in the SI. As expected, the samples featuring a single overhang approximately in the centre of the construct (ssx1 and hyx1) displayed a sub-event current spike approximately in the centre of the respective event (see section 3 of the SI for a discussion on DNA knotting). The position of the spike is thus invariant with regards to the orientation the DNA enters the pore in. For the samples with two overhangs, ssx2 and hyx2, the situation is slightly more complex in that the second, off-centred overhang can appear before or after the central one, depending on which part of the DNA carrier enter the pore first. This is illustrated in the examples given for ssx2 (bottom left), where the off-centred overhang appears after the centred one in event 1 and 3 (top and bottom), and before in event 2 (middle). These considerations are also borne out in the statistical analysis of the sub-event positions, as shown in panel B) for ssx1 and hyx1 (top) and ssx2 and hyx2 (bottom). There are however several observations worth noting: First, all events shown in panel A) share some common features, in terms of their overall shape. Namely, they all start with a relatively sharp $I(t)$ transition as the DNA enters the pore channel from the outside, reflecting the relatively abrupt boundary between the pore entrance and the bulk solution. The current level then remains relatively constant until there is first an abrupt change and a non-linear tail-off. Again, this most likely reflects the internal geometry of the pore channel and, in particular during tail-off, how the DNA leaves the narrowest part of the channel towards the bulk solution inside the pipette. With geometrically simple and well-defined analytes, such as spherical nanoparticles, this effect has previously been exploited to reconstruct the inner shape of the pore channel.[46] Secondly, the normalised histograms of the sub-event positions for ssx1 vs. hyx1 and of ssx2 vs. hyx2 strongly overlap, suggesting that the hybridisation state of the overhang has little effect on the translocation characteristics of the carrier DNA. This is despite the fact that the pore diameter is smaller than the length of the overhangs, which are in turn shorter than the persistence length of double-stranded DNA (>35 nm).[47,48] The same conclusion is however also borne out in more detailed analysis of the translocation events below. Finally, the peak positions for all four samples are in excellent agreement with expectations based on the DNA design and in accordance with the AFM data above. From Gaussian fitting, ssx1 and hyx1 feature a single peak at relative positions of 0.51 ± 0.04 and 0.53 ± 0.07 (normalised to $\tau_e$, error: ± 1σ). The expected value based on the DNA design is 3576/7228 = 0.49 or (7228 - 3576)/7228 = 0.51, depending on the DNA orientation, a difference that is within experimental error. For comparison, the AFM characterisation of hyx1 yielded a relative overhang position of 0.51 ± 0.05, *vide supra*. For ssx2, peaks occur at 0.32 ± 0.04, 0.50 ± 0.05 and 0.69 ± 0.05, those for hyx2 at 0.33 ± 0.06, 0.50 ± 0.05, 0.68 ± 0.04 (fit: sum of 3 Gaussians). The expected values are 0.32 and 0.51 for the two overhangs in one translocation direction, and 0.49 and 0.68 for the other. The combined peak positions are again in very good agreement with the experimental values, within error. We also note that the observed intensity ratio is approximately 1:2:1, which is expected if the two DNA ends enter the pore with roughly equal probability (actual values, from triple Gaussian fits, ssx2: 1:2.1:1.3; hyx2: 1:1.8:1.4). Taken together, these data strongly suggest that the preparation of the DNA designs has been successful in all four cases.

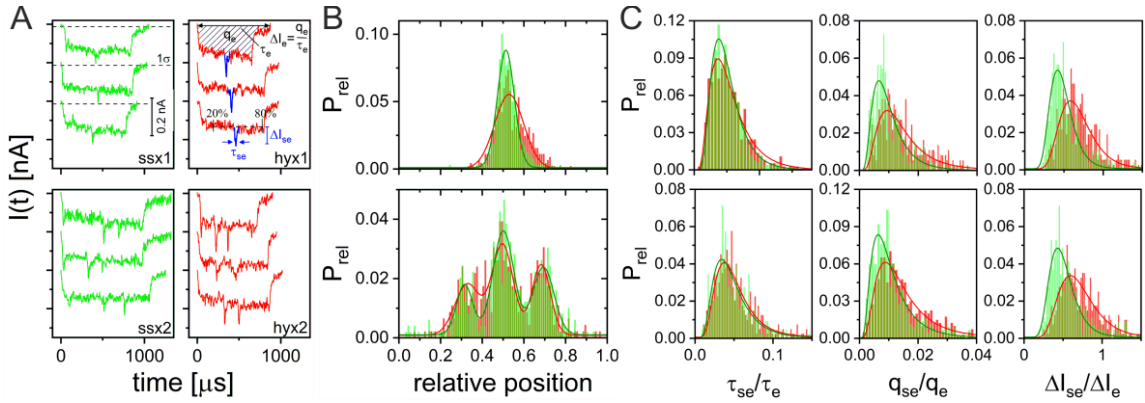

**Fig. 4**: DNA carrier translocation data, analysis of sub-events ($V_{bias}$ = 0.7 V). A) Example events for the four DNA structures analysed here: ssx1/hyx1 (top) and ssx2/hyx2 (bottom). The color-coding, ssx (green) and hyx (red), is the same throughout this figure. Key parameters characterising the events and sub-events are illustrated, including the 1σ line defining event start/stop according to our definition and the 20%/80% boundary for the sub-event search (see Methods). Some sub-events, as defined by the search algorithm used here, are shown in blue. B) Normalised histograms of the relative sub-event position for single-overhang samples (ssx1, hyx1; top) and the double-overhang samples (ssx2, hyx2; bottom), inc. Gaussian fits. C) Normalised histograms of sub-event characteristics, relative to the respective event: $\tau_{se}/\tau_e$, $q_{se}/q_e$ and $\Delta I_{se}/\Delta I_e$. All distributions are non-Gaussian and are represented well by log-normal fits (solid lines). Importantly, as the aim of the study is to distinguish hybridised (red) from non-hybridised overhangs (green), the difference between the two cases appears to be largest for the $\Delta I_{se}/\Delta I_e$ distributions (This is also the case for the data recorded at $V_{bias}$ = 0.5 V, see SI).

We now address the key question of the present study, namely whether the hybridisation state of the overhangs can reliably be determined using resistive pulse sensing under the present conditions. Three fundamental signal properties were explored in this context, namely the sub-event duration $\tau_{se}$, the sub-event charge $q_{se}$ and the maximum current within a sub-event, $\Delta I_{se}$. This was based on the consideration that a stiffer (hybridised) overhang may increase the residence time in the sensing zone (and hence $\tau_{se}$), and that the increased presence of DNA could increase $q_{se}$ or enhance blockage (thus $\Delta I_{se}$). Due to the relatively large variance in each of these event characteristics, we found it necessary to normalise $\tau_{se}$, $q_{se}$ and $\Delta I_{se}$, to the corresponding event properties for each event. Since in our DNA design the overhangs are either single-stranded or double-stranded, there is no obvious internal reference for this normalisation process, in contrast to design used by others.[22] Accordingly, the respective normalised histograms for all three cases, $\tau_{se}/\tau_e$, $q_{se}/q_e$ and $\Delta I_{se}/\Delta I_e$, are shown in fig. 4 C) (top: ssx1/hyx1, bottom: ssx2/hyx2, in green and red, respectively). The solid lines are fits to log-normal distributions, which generally provide a very good representation of the histograms. The individual values of the fit parameters are of less relevance here, but the fact that the data are not normally distributed affects the statistical analysis, as discussed below. From the histogram shapes it is apparent that of those three classification parameters, the $\Delta I_{se}/\Delta I_e$ histograms show the largest differences between the ssx and hyx samples. The same observations hold true for the data recorded with a different set of pipettes at $V_{bias}$ = 0.5 V, as shown in the SI (section 5).

In order to test whether the observed differences in the $\Delta I_{se}/\Delta I_e$ distributions were statistically significant, we subsequently performed a three-factor Analysis of Variance (ANOVA) with two levels each, taking into account the hybridisation state of the overhangs (ssx vs. hyx), the bias voltage (0.5 V vs. 0.7 V) and the number of overhangs per carrier ('single' vs. 'double'). In this context, it is worth re-iterating that the overhangs in all samples have the same sequence composition. Accordingly, our analysis initially considers whether there is any significant difference between ssx and hyx, irrespective of the sample and conditions used. An ANOVA is thus the preferred method, as it allows for multiple comparisons to be performed at the same time (different factors, such as experimental conditions and samples, for example) and provides information on the interactions between those factors. It is more conservative than performing multiple t-tests and avoids an accumulation of type-I errors (false positives).[49] In a second step, we also performed a series of *post hoc* Tukey-Kramer 'means difference' tests for the individual comparisons (ssx1 vs. hyx1, ssx2 vs. hyx2 at the two different voltages) to investigate the observed main effects and some of the interactions in more detail.

Conventional ANOVA has several requirements, namely that the data be (approximately) normally distributed, possess equal (or comparable) variance and be independent. The latter is the case here, since the individual datasets were recorded in separate experiments in several (23) different pipettes. ANOVA is relatively robust with regards to data that are not normally distributed, but in light of the results shown in fig. 4 C) (log-normally distributed data), the $\Delta I_{se}/\Delta I_e$ data were log-transformed first. A Kolmogorov-Smirnov test was then performed to check for normality (cf. SI). In all eight cases (datasets for ssx1, hyx1, ssx2 and hyx2 at $V_{bias}$ = 0.5 V and 0.7 V, respectively) the distributions appeared close to normal, only for one dataset was normality formally rejected at a confidence level of 0.05 (ssx1 at $V_{bias}$ = 0.7 V, p = 0.013). A Levène test was performed to compare the variances of the different distributions, which were found to be not significantly different at a confidence level of 0.05. As a result, we felt the requirements for an ANOVA were sufficiently fulfilled to proceed with the analysis.

The results of the three-factor ANOVA are shown and discussed in detail in sections 6 and 7 of the SI (including interaction effects). Focusing on the main effects, the means of two factors, namely 'hybridisation state' and 'voltage', are statistically significantly different at a 0.05 confidence level (p ≈ 0 and 0.008; means difference: -0.139 and 0.021; sample sizes: 1266 (ssx)/1088 (hyx) and 657 (0.5

V)/1697 (0.7 V)), while the third factor, 'number of overhangs', is not (p = 0.061; means difference: 0.015; sample size: 1224 (single overhang)/1130 (double overhang)). We note that the sample size in the 'voltage' comparison is unequal, which however does not seem to affect the homogeneity of variance, as shown above. This is thus unlikely to compromise the conclusions from the ANOVA. The effect of 'hybridisation state' is clearly relevant to the underlying idea of the present work and will be explored in further detail below. A main effect of 'voltage' could also be of interest in that it could suggest that optimisation of $V_{bias}$ could lead to improved sensor performance. However, as discussed in section 7 of the SI, when considering interaction effects and suitable *post hoc* tests, the effect was found not to be statistically robust, in terms of the individual comparisons (i.e., differentiation between ssxx and hyx was similar at the two $V_{bias}$ values studied). Finally, the absence of statistically significant effects for the number of overhangs would suggest that the latter does not affect the detection of the individual sub-events, at least in the present samples and condition used.

To confirm whether the hybridised and non-hybridised overhangs in a given sample could indeed be differentiated in a statistically significant manner, we performed Tukey-Kramer 'means difference' tests for each of the four relevant individual comparisons, *cf*. fig. 5 (Panels A-D: ssx1 vs. hyx1, ssx2 vs. hyx2, at $V_{bias}$ = 0.5 V and 0.7 V, respectively). Indeed, in all cases, the difference was found to be statistically significant at $\alpha$ = 0.05 (type-I error rate). Moreover, in all four cases, the means difference between the ssx and hyx samples have the same sign (A to D: -0.095; -0.199; -0.133; -0.129) and are on average of similar magnitude (i.e., there is no obvious difference between the single- and double-overhang samples, in accordance with the discussion above). Thus, the $\varDelta I_{se}/\varDelta I_e$ ratio is slightly, but consistently larger for hybridised overhangs, compared to their single-stranded analogs.

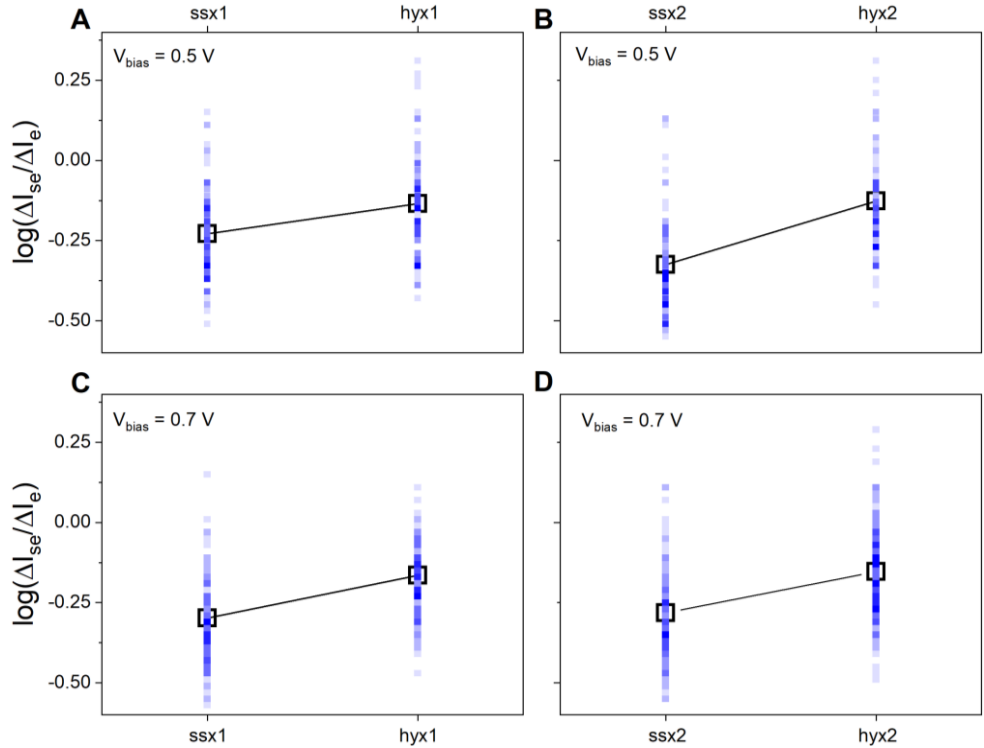

**Fig. 5**: Result of the pair-wise comparisons ssx1 vs. hyx1 and ssx2 vs. hyx2 at $V_{bias}$ = 0.5 V (A/B) and 0.7 V (C/D). The respective mean values (black squares) and individual data points (density maps, blue/white: high/low density) are shown. The means difference is comparable in magnitude in each case and such that $\log(\Delta I_{se}/\Delta I_e)$ is consistently larger for hyx samples, compared to the corresponding ssx samples (see main text for further discussion).

In summary, our results confirm that the hybridisation state of the overhangs may be detected in a statistically significant manner, based on a sufficient number of translocation events. The signal-to-noise ratio, i.e. the difference between ssx and hyx samples, is relatively small at present, but may be improved further. $V_{bias}$ is a parameter we considered in this context, but in light of the analysis above, we did not find a strong dependence on $V_{bias}$. Moreover, the accessible $V_{bias}$ range is relatively limited: At small $V_{bias}$, the translocation frequency becomes very low and the time required to record a sufficient number of translocation events very long. This is not practical for real-life sensor applications. At too high $V_{bias}$, translocation becomes very fast and the sub-events are increasingly difficult to resolve.

Decreasing the pore size may be another strategy towards improving the sensor performance and is known to improve the signal-to-noise ratio for translocation events.[44] Atomic Layer Deposition (ALD) of oxides has been explored in this context, also in nanopipettes.[50,51] However, care needs to be taken with regards to the identity and surface properties of the oxide as well as the preparation conditions. For example, we observed that some $Al_2O_3$ films were not stable under the experimental conditions used in the present study, i.e. at very high halide concentrations, in line with previous literature.[52,53,54,55] Moreover, in the presence of the $Al_2O_3$ layer, the channel surface is net positively charged,[50] which attracts the negatively charged DNA. This leads to adsorption of the DNA to the pore surface, wider translocation time distributions and less well resolved individual events. Hence, a different oxide, for example with negative surface charge in solution, may be preferable in this regard.

**Conclusions**

In conclusion, we have demonstrated how nanopipette-based, all-electric detection, combined with robust statistical analysis, is capable of probing the hybridisation state of short, approximately 100 nt long single-stranded overhangs. Their length is comparable with short circulating DNA fragments, which are found in different bodily fluids and have been identified as potential markers in disease diagnostics, for example in TB detection. Accordingly, and to illustrate this aspect, the 88 nt probe design employed in this study was taken from the RV1910c gene, a gene regulatory region of the *KatG* protein known to play a key role in TB resistance against INH, one of the most effective and specific anti-TB drugs available. With the overhangs arranged over a long DNA carrier, the sensing strategy encompasses significant multiplexing capabilities. Not only is it possible to integrate a larger number of probes in one carrier (of equal or different sequence composition), but also to mix carriers of different lengths (and different overhangs). Equally, it would also be possible to encode specific features into the carriers (such as hairpins), to differentiate those of equal length.[22] However, while the carrier designs used in present study have been implemented using DNA self-assembly from small fragments, similar structures may be created employing enzymatic modification.[56,57,58,37] The latter may be significantly more cost-effective and enable the preparation of larger amounts, both required for a viable sensor technology in the future. The incorporation of several larger features on the other hand may complicate the preparation of the carriers, which would be less desirable. Following first efforts to improve the sensor performance, we have identified a small number of parameters to be explored in this context in the future. Finally, and most importantly, our study demonstrates the remarkable sensitivity of electric, nanopipette-based sensing towards the detection of even minor changes in DNA structure or composition, in a fast and label-free manner.

**Experimental Section**

*Preparation and characterisation of the DNA constructs.* The synthesis was adapted from Bell et al. and Plesa *et al*.[22,23] The restriction enzymes (RE) (BamHI-HF, EcoRI-RF), M13mp18 circular ssDNA and M13mp18 RF circular dsDNA were purchased from New England Biolabs (NEB) (Ipswich, U.K.). The 190 staples, ssDNA overhangs sequences, and target DNA sequence were purchased from Integrated DNA

Technologies (IDT) (Leuven, Belgium). The sequence for the staples were similar to the ones reported by Bell et al.[19] These overhang and target strand sequences can be found in the SI.

The M13mp18 ssDNA was linearised to form the DNA carrier strand. The oligonucleotides with the sequence 5' – TCT AGA GGA TCC CCG GGT ACC GAG CTC GAA TTC GTA ATC – 3' were hybridized to the ssDNA to form a double stranded restriction site recognisable by the RE. For the hybridization and RE cutting, 5 µL of M13mp18 (250 ng/µL), 5 µL of 10x NEB 'cut smart' buffer, 1 µL of oligonucleotide (100 µM) and 37 µL of autoclaved ultrapure water was mixed. To hybridise the oligonucleotide, the mixture was heated to 65 °C for 5 minutes, followed by cooling at 25 °C for 5 min. and further cooling at 10 °C for 10 min. in a thermocycler (Eppendorf Mastercycler Gradient).

After which, 1 µL of BamHI-HF and 1 µl of EcoRI-HF (20,000 units/mL) were added to the mixture. The mixture was incubated in the thermocycler at 37 °C for 2.5 hours and then heated to 65 °C for 20 min. to denature the RE. The ssDNA was cleaned up using the Monarch PCR & DNA Clean-up kit (NEB, Ipswich, U.K.) and eluted in autoclaved TE buffer (10 mM Tris-HCl, 1 mM EDTA, pH 7.8, Sigma Aldrich, Irvine, U.K.). A purity check (running the eluted DNA on an 0.8 % agarose gel) was performed after each digestion. The final concentration of the ssDNA was measured with UV-visible spectroscopy (Nanodrop 1000, Thermo Scientific). The same procedure (without hybridisation) was also used to form the linearised dsM13mp18 from the circular M13mp18-RF.

To form the dsDNA carrier strand with one double stranded overhang, 42 µL of linearised M13mp18 (11 nM), 1 µL of staple mix (38 bp oligonucleotides, each oligo 30 µM), 2 µL of both overhang strands (short and long) (100 µM each), 2 µL of target sequence (100 µM), 5 µL of $MgCl_2$ (100 mM) and 8 µL of autoclaved ultrapure water were mixed. The mixture was heated to 72 °C and cooled at a rate of 1 °C every 4 minutes till the temperature dropped to 23 °C. The excess staples, overhangs and target strands were (partially) removed using the Amicon Ultra 100 kDa cut-off centrifugal filters (Millipore, Massachusetts, U.S.). The purification step consisted of diluting the mixture with 400 µL of TE buffer and then centrifuging at 3000 g for 10 min. at 4 °C. The filtrate was then decanted and the above procedure repeated for 6 washing steps. The sample was recovered by inverting the filter and centrifuging at 1000 g for 2 min. The product quality and quantity were initially characterised by gel electrophoresis and UV-vis spectroscopy. A similar procedure was carried out to form constructs with a single (unhybridised) ss-overhang and those with two overhangs.

*AFM studies.* The DNA samples were imaged in tapping mode in air at 23 °C with an Agilent 5500 AFM/SPM microscope (Keysight Technologies, Arizona, U.S.A.) and commercial "PointProbe® Plus-NCHR-10" probes (Windsor Scientific, Slough, U.K.). Images were processed with the 'plane' and 'flatten' filters in the WSxM 5.0 Develop 7.0 software.[59] Preparation of the substrate: the buffer (10 mM HEPES, pH 7.6, 4 mM $MgCl_2$, 1 mM EDTA; Sigma Aldrich) was filtered using a 0.2 µm syringe filter (EMD Millipore Hertfordshire, U.K.) to remove any large particle contaminants and then autoclaved. The construct (1.5 ng/μL) was prepared in 20 μL of the said buffer and deposited on a freshly cleaved Mica (9.9 mm diameter, Agar Scientific, Stanstead, U.K.). The DNA was left to adsorb to the surface for 5-10 min. The surface was then rinsed with 1 mL of nuclease-free water (NEB) thrice and dried in a $N_2$ gas flow.

*Nanopipette fabrication and characterisation.* Nanopipettes were made from filamented quartz capillaries (O.D.:1 mm, I.D.:0.5 mm; length: 7.5 cm, Sutter Instruments, Novato, USA). The capillaries contain a ~160 μm glass filament that facilitates the filling of the nanopipette with electrolyte by capillary action.[60] The glass capillaries were first plasma-cleaned for 7 minutes (Harrick Scientific, New York, U.S.A.) before loading it into the laser pipette puller (P2000, Sutter Instruments). The pulling programme involved two steps and the following parameter settings: Step 1 ('Heat': 880-890; 'Filament': 4; 'Velocity': 30; 'Delay': 175-190; 'Pull': 100-110). Step 2 ('Heat': 900; 'Filament': 1; 'Velocity': 15-20; 'Delay': 170-175; 'Pull': 160). Minor re-optimisation of the parameter settings was sometimes required, when the apparent pore diameters consistently veered off the desired range from 20-30 nm (as judged by the pore conductance, *vide infra*). This was most likely due to small changes in the environmental conditions (humidity, ambient temperature) or the puller itself, but those changes were small, as shown above. The inner diameter of the nanopore at the end of the pipette was initially estimated from the conductance of the pipette (see SI) in 1 M KCl and, in some cases, characterised further using TEM. For translocation experiments, the pipettes were then integrated into a custom-built liquid cell, with one Ag/AgCl electrode on the inside and the other one on the outside of the pipette.

*TEM characterisation of the pipettes*. TEM imaging of the nanopipettes was carried out using the JEOL JEM-2100F TEM. The measurements of the images were conducted using Image J.[61] Sample preparation: The tip of the pipette is positioned such that it was sitting parallel on the centre of the Cu TEM slot grid (Cat no.:GG030, Taab Laboratory Equipment Ltd, Aldermaston, U.K.) and glued to the grid with a two-component epoxy glue (Araldite, Basel, Switzerland). The glue was left to set for 6 hours, after which the pipette attached to the grid was cleaned under UV/Ozone for 20 mins (UVOCS). It was then sputter coated (Polaron Quorum Technologies, East Sussex, U.K.) with 10 nm Cr to reduce charging effects. The parts of the pipette lying just outside the grid was cut off using a scalpel before the grid was placed in the sample holder of the TEM.

*Translocation experiments* were performed in 4 M LiCl electrolyte, which is known to reduce the translocation speed in comparison to KCl,[62,63] using a custom-made low-noise, wide-bandwidth current amplifier reported previously.[16,17] The electronics output is split into a 'DC' and an 'AC' channel, containing slow (below 10 Hz) and fast (> 10 Hz) modulations of the current, respectively.

Specifically, this means that translocation events appear in the AC output. It is zero mean, which greatly simplifies any background correction (minor constant offsets were corrected prior to the event search, *vide infra*). The DC channel contains the steady-state current through the cell, which is related to the pore conductance. The AC output is filtered as specified with an eight-pole low-pass (analog) Bessel filter (Krohn-Hite Corporation, Massachusetts, U.S.A.). A digital oscilloscope (Picoscope 4262, Pico Technology, Cambridgeshire, U.K.) served as analog-to-digital converter at 1 µs sampling rate. Custom-written Matlab code was used for instrument control, data acquisition and analysis, as detailed below. The liquid cell and the amplifier were housed in a double-Faraday cage to minimize electrical noise. In total 23 different nanopipettes were used in the translocation experiments presented in this work.

*Analysis of the translocation data.* Current-time traces were initially subjected to a zero-order background correction to account for minor, constant offsets in the AC channel output (typically < 10 pA). Then, a threshold search was performed with a 5σ cutoff, where σ is the standard deviation of the current noise in the AC channel. The search algorithm then found the data points, which first crossed the zero baseline, relative to the 5σ cutoff, as well as the corresponding 1σ values. The latter served as definition for the event start and stop, as a compromise between minimizing the effect of local baseline fluctuation on the event characteristics and our ambition to capture the overall event shape as much as possible. Thus, the '1σ' event duration is $\tau_e = t_{stop} - t_{start}$. The probability distribution of $\tau_e$ was found to be skewed and well approximated heuristically by a log-normal distribution.[16,17] For a physically rigorous closed-form solution of the distribution function, see reference 45. The effective event magnitude $\Delta I_e$ was calculated from the integral of the *I*(*t*) trace between $t_{start}$ and $t_{stop}$, $q_e$, divided by $\tau_e$.

Sub-events are more challenging to identify and analyse, because they typically contain far fewer data points than the events themselves. We therefore took a somewhat different approach in searching for and analysing those sub-events. First, the median of a central section of the event (0.2 to 0.8 relative event duration) was determined to serve as baseline (the median, rather than the mean, was chosen to be less sensitive to outliers, such as spikes). Sub-events were identified in a threshold search with a 1.2·$\Delta I_e$ cutoff. The search algorithm also extracted all adjacent data points before and after, until the median value was reached. The sub-event duration $\tau_{se}$ is thus the time difference between the first median crossing after the sub-event threshold was reached and the last median crossing before the threshold value (capturing a large part of the sub-event shape). $\Delta I_{se}$ was taken to be the maximum current, relative to median, and $q_{se}$ was determined from the integral of the *I*(*t*) trace, within a sub-event. In absence of a closed-form solution for the distribution functions of $\Delta I_{se}/\Delta I_e$, $q_{se}/q_e$ and $\tau_{se}/\tau_e$, the corresponding probability distributions were also approximated by log-normal distributions.

These were mainly used for illustration purposes and to highlight the non-normality of the data, in the context of the subsequent statistical analysis.

**Acknowledgments**

The authors thank Willem van Schaik (Institute for Global Innovation and Institute of Microbiology and Infection, University of Birmingham, UK), Robert Neely (School of Chemistry, University of Birmingham) and Ekaterina Kornysheva (Department of Psychology, Bangor University, UK) for discussions.

**Author contributions**

T.A. conceived the experiment. T.A., A.Y.Y. Loh and C.H Burgess performed the experiments and analysed the data. The manuscript was written with contributions from all authors.

**Competing Interests**

The authors declare no competing interests.

**Data availability**

The datasets generated during and/or analysed during the current study are available from the corresponding author on reasonable request.

**Code availability**

Scripts for the analysis of the translocation data presented in this manuscript are available from the corresponding author on reasonable request.

# An All-Electric Single-Molecule Hybridisation Detector for short DNA Fragments

A.Y.Y. Loh,[1] C.H. Burgess,[2] D.A. Tanase,[1] G. Ferrari,[3] M. Maclachlan,[2] A. Cass,[1] T. Albrecht*[1,4]

[1]Imperial College London, Department of Chemistry, Exhibition Road, London SW7 2AZ, UK

[2]Imperial College London, Department of Materials and Centre for Plastic Electronics, London SW7 2AZ, United Kingdom

[3] Politecnico di Milano, Dipartimento di Elettronica, Informazione e Bioingegneria, P.za Leonardo da Vinci 32, Milano, Italy

[4]University of Birmingham, School of Chemistry, Edgbaston Campus, Birmingham B15 2TT, UK

*t.albrecht@bham.ac.uk

## 1) DNA design and characterisation

DNA sequences used in the overhang region and illustration of the design:

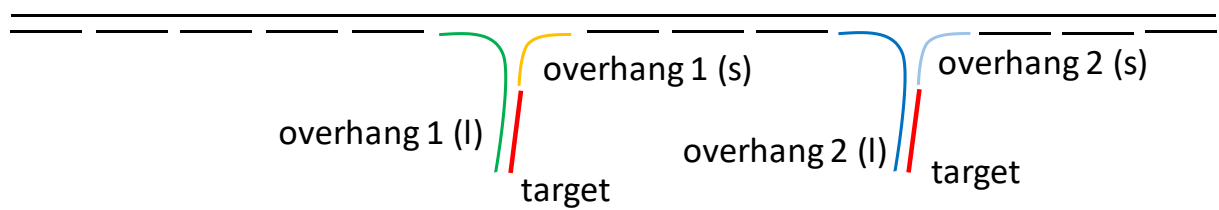

**Fig. S1**: Illustration of the DNA design with focus on the overhang regions, for ssx2/hyx2 (ssx1/hyx1 only contain overhang 1). The overall length of the overhangs is 100 nt, where the 12 nt closest to the carrier are hybridised (i.e., between overhang 1 (l) and overhang 1 (s), and overhang 2 (l) and overhang 2(s), respectively) to form a 'stalk' and stabilise the region. The remaining 88 nt of overhangs 1 (l) and 2 (l) act as probe and are complementary to the target.

**Overhang 1 (s)**: 5' - ACT CCG ACC GAG CGC TGC TGC TTT CGG CGC CAG TAG CAC CAT TAC CAT TAG CAA GGC CGG AAA CGT CAC C – 3'; **overhang 1 (l)**: 5' - CTT GAG CCA TTT GGG AAT TAG AGC CAG CAA AAT CAC CAT GGC GCC GAA AGC AGC AGC GCT CGG TCG GAG TAT GCC CGA AAC GCC TAC CGG CGA TGT ACT GAC AAT CAG CAG TCC GGC ATT CGC CGA CGG TGC GCC GAT CCC GGA ACA GTA CAC CTG CA– 3'; **overhang 2 (s)**: 5' - TCG CTG GCA GCG TAC CGC GGC GGT CTG AGC CGT AGT GGC AAA TCC AAT CGC AAG ACA AAG AAC GCG AGA A – 3'; **overhang 2 (l)**: 5' - CTC GGG CTT AGG TTG GGT TAT ATA ACT ATA TGT CAC TAC GGC TCA GAC CGC CGC GGT ACG CTG CCA GCG AAT GCC CGA AAC GCC TAC CGG CGA TGT ACT GAC AAT CAG CAG TCC GGC ATT CGC CGA CGG TGC GCC GAT CCC GGA ACA GTA CAC CTG CA– 3'; **target**: 5' - TGC AGG TGT ACT GTT CCG GGA TCG GCG CAC CGT CGG CGA ATG CCG GAC TGC TGA TTG TCA GTA CAT CGC CGG TAG GCG TTT CGG GCA T– 3'

Thermodynamics of binding between overhang and target strands:

In order to confirm that the target sequence indeed bound to the probe with high enough efficiency, we calculated the concentration conditions based on thermodynamic, equilibrium binding considerations for the ssx1/hyx1 samples. For this purpose, we used the on-line implementation of OligoCalc with the 88 nt target sequence given above,[1] determined the Gibbs free energy of binding,

$$\Delta G = R \cdot T \cdot \ln\left(\frac{[hyx1]}{[ssx1]\cdot[target]}\right) = 634 \frac{kJ}{mol} \tag{S1}$$

and finally calculated the solution concentrations of the relevant species (T = 298 K), fig. S2.

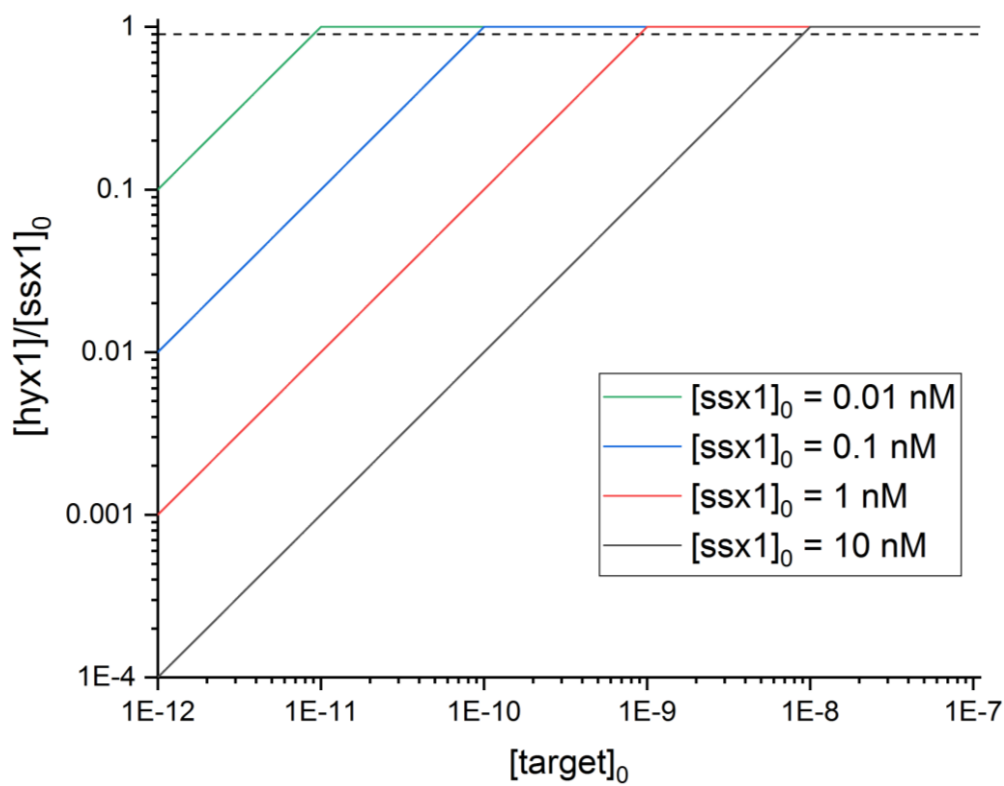

**Fig. S2**: The concentration of hyx1, [hyx1], relative to the initial concentration of ssx1, $[ssx1]_0$, as a function of the initial target concentration, $[target]_0$. $[ssx1]_0$ was varied, as indicated. Dashed line: $[hyx1]/[ssx1]_0 = 0.9$. Calculations were performed using eq. S1b, which has been re-arranged from eq. S1 with x = [hyx1], [target] = $[target]_0$ - x and so forth. The plot illustrates that the limit of detection, in terms of $[target]_0$, may be adjusted via $[ssx1]_0$. Considerations for ssx2 and hyx2 are similar, bearing in mind that there are now two, most likely independent binding sites.

$$[hyx1] = \frac{A \cdot ([ssx]_0 + [target]_0) - B + 1}{2A} \tag{S1b}$$

where $A = \exp(\Delta G/(RT))$ and $B = \sqrt{A^2 \cdot ([ssx]_0 - [target]_0)^2 + 2A \cdot ([ssx]_0 - [target]_0) + 1}$.

Gel electrophoresis results for some of the carrier constructs and other samples, as described:

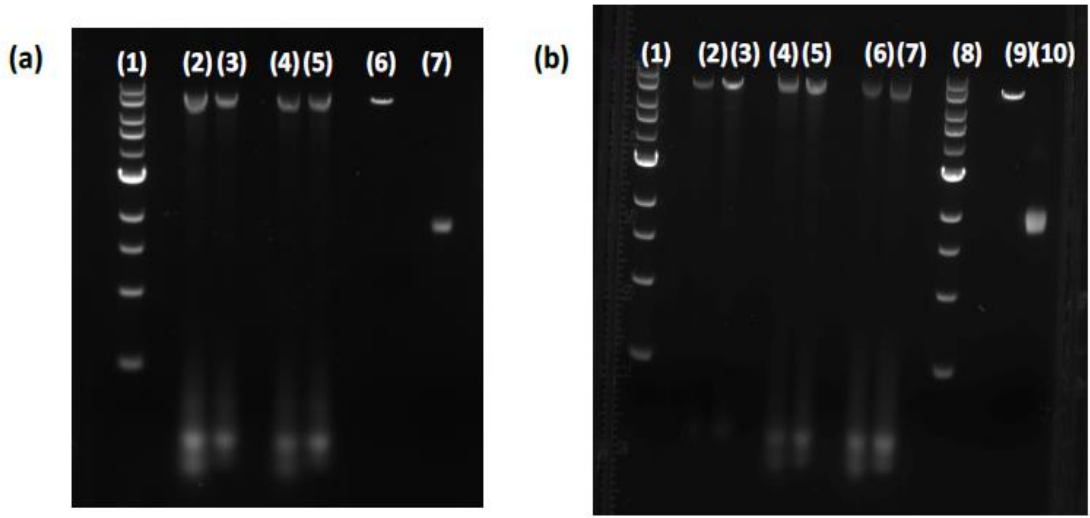

**Fig. S3:** Gel shift assays of the carrier constructs and other samples as specified. a) Row 1: 1 kb ladder; 2/4: hyx1 before purification; 3/5: hyx1 after purification; 6: linearised ds M13mp18 DNA; 7: circular ss M13mp18 DNA. b) Row 1: 1 kb ladder; 2/3: ssx1 before/after purification; 4/5: ssx2 before/after purification; 6/7: hyx2 before/after purification; 8: 1 kb ladder; 9: linearised ds M13mp18 DNA; 10: circular ss M13mp18 DNA. Purification was performed with Amicon Ultra 100 kDa cut-off centrifugal filters, as described in the main text.

## 2) Conductance measurements and an estimation of the sensing zone

Electrode preparation: Ag wire (diameter: 0.25 mm; length: 7.1 cm, Goodfellow Cambridge Ltd, UK) was first cleaned by immersion in 34 % nitric acid (VWR International, Pennsylvania, USA) for 15 s. AgCl was deposited in 1 M KCl electrolyte (VWR International) with chronopotentiometry ($I$ = 0.5 mA for 500 s). The Ag/AgCl electrodes were then soldered to gold pins to connect to the nanopore translocation set-up. The electrolyte solution was either 1 M KCl with TE buffer (10 mM Tris-HCl, 1 mM EDTA, pH 7.8, Sigma Aldrich) or 4 M LiCl with 10 mM TE, as specified. The electrolytes were filtered using a 0.2 µm syringe filter (EMD Millipore) to remove any large particle contaminants and then autoclaved. The nanopipette was back-filled with the electrolyte using a syringe needle (MicroFil, World Precision Instruments, Florida, U.S.A.) attached to a 1 mL syringe (NormJect Luer, Henke Sass Wolf, Germany). Air bubbles trapped in the nanopipette tip were removed by rasping with the corrugated end of a pair of tweezers. Glass vials used to contain the electrolyte (3 mL) were first sonicated in EtOH (VWR International) (2 rounds, 10 min. each) and then in ultrapure $H_2O$ (3 rounds, 10 min. each) before leaving them to dry in an oven. They were then autoclaved for sterilisation prior to use.

The conductance of the nanopipette was determined from the slope of the $I/V$ trace at low voltage in electrolyte (1 M KCl, 10 mM Tris-HCl, 1 mM EDTA or 4 M LiCl, 10 mM Tris-HCl, 1 mM EDTA) using cyclic voltammetry (Gamry Reference 600 Potentiostat, scan rate: 0.1 V/s, 2-electrode configuration), recorded in a potential range from +0.5 V to -0.5 V. The inner diameter $d_p$ of the pipette tip was then estimated using the following equation:[2]

$$d_p = \frac{4GL + \frac{\pi}{2}GD_i}{\pi D_i g - \frac{\pi}{2}G} \tag{S2}$$

where $G$ is the conductance and $L$ the taper length of the nanopipette, $D_i$ the inner diameter of the capillary (0.5 mm in our case), and $g$ is the conductivity of the electrolyte (as measured with conductivity meter (Mettler-Toledo, Greifensee, Switzerland). The taper lengths of all the pipettes used were measured by optical microscopy (Olympus, Tokyo, Japan) from the tip of the capillary to where the internal channel had reached $D_i$. The values obtained corresponded rather well to those obtained from SEM and TEM imaging, which was performed on a small selected number of pipettes. As discussed in the main text, the channel geometry is close to conical even rather far along the pipette and away from the tip. Since the pore diameter increases accordingly, the contribution to the overall resistance of the decreases markedly. The resistance of a conical pipette is given by

$$R = \frac{\rho}{\pi} \cdot \frac{L_c}{r_{in} \cdot r_{out}} \tag{S3}$$

where ρ is the resistivity of the solution, ρ = 1/g; $L_c$ the axial length of the conical channel; and $r_{in}$ and $r_{out}$ the pore radii at the entrance and exit of the channel. It can then be shown the resistance of a conical pore of length $L_c'$, relative to one of length $L_c$, is (with $L_c' \leq L_c$):

$$\frac{R(L_c')}{R(L_c)} = \frac{L_c' \cdot (L_c \cdot tan(\alpha) + r_{in})}{L_c \cdot (L_c' \cdot tan(\alpha) + r_{in})} \tag{S4}$$

$\alpha$ is the opening angle of the cone, relative to the pore axis.

This relation is plotted in fig. S4 below. It demonstrates that even within the first 10 nm from the pore entrance, almost 20% of the resistance drop occur. After 100 nm, this resistance drop already exceeds 70%. Thus, the sensing zone, i.e. the region that is most sensitive to changes in resistance lies within the first few 10s of nanometers from the pipette tip.

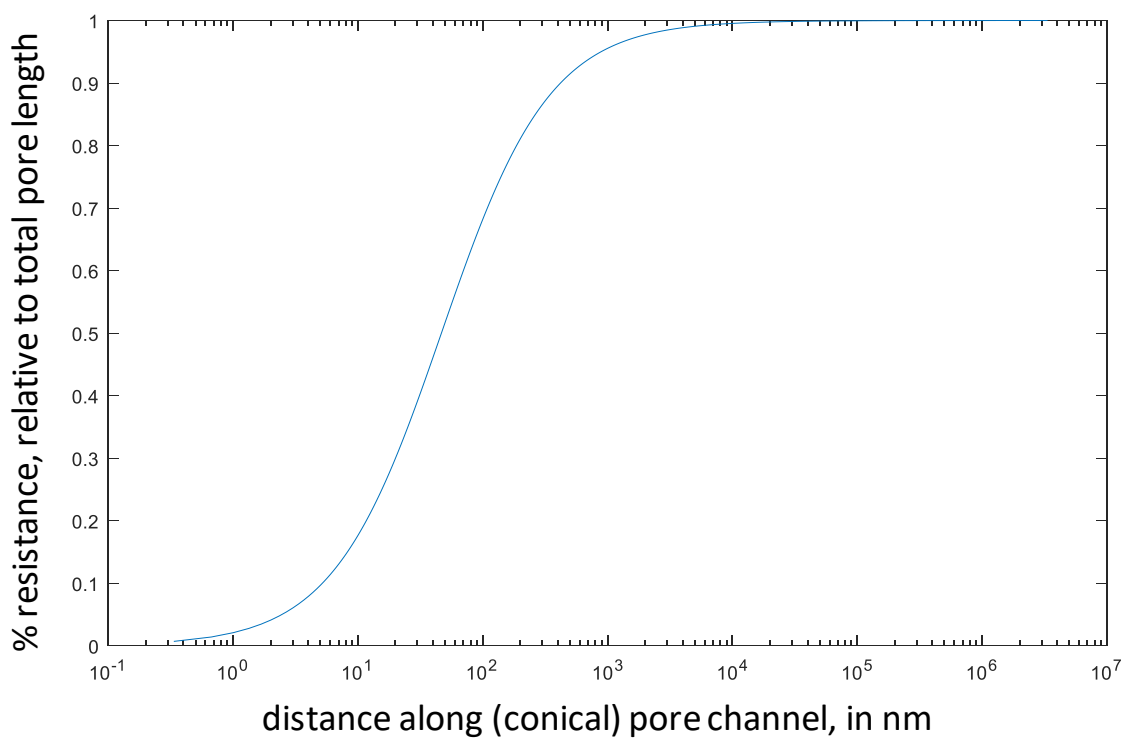

**Fig. S4**: Relative resistance of a conical pore as a function of $L_c'$, at given $L_c$, according to eq. S4. Opening angle $\alpha$ = 15°; membrane thickness $L_c$ = 3.35 mm; pore radius, entry side: $r_{in}$=12.5 nm. The figure illustrates the relative contribution to the overall resistance of the pore, as the distance $L_c'$ along the pore channel is increased (counted from the pore entrance).

## 3) DNA knotting

It has been reported previously that even translocation of bare double-stranded DNA can produce current-time transients including a 'spike-like' feature, for example due to knotting.[3] Spikes resulting from the latter are typically roughly three times larger than the current level associated with linear ds DNA, due to the geometry of the knot. Since similar spike features are an important aspect in our study as well, we wanted to understand whether they occur and how abundant they are in our case, ultimately to rule out artefacts in the interpretation of our core results. For this purpose, translocation experiments were performed with 7.2 kbp (ds) M13mp18 DNA that did not contain any overhangs. Indeed, some events resembled the pattern expected for DNA knots, as shown in fig. S5. It turned out, however, that they were rather rare and occurred in only 3.1% of 820 translocation events, in line with the previous literature (4.4 % in ref. 3). We therefore felt that their presence would not seriously affect the experiments with the overhang-containing samples and did not make any attempts to separate them out in those datasets.

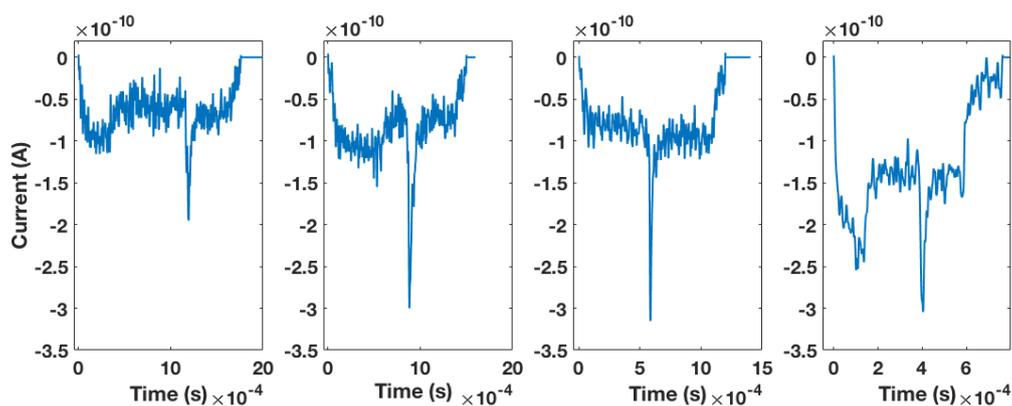

**Fig. S5**: Example events from the translocation of 7.2 kbp M13mp18 DNA at different $V_{bias}$ (0.3 V; 0.5 V; 0.5 V and 0.7 V from left to right (4 M LiCl electrolyte; $d_p$ = 19 nm; 100 kHz filter frequency)

## 4) Comparison of translocation data for hyx1 and ssx1

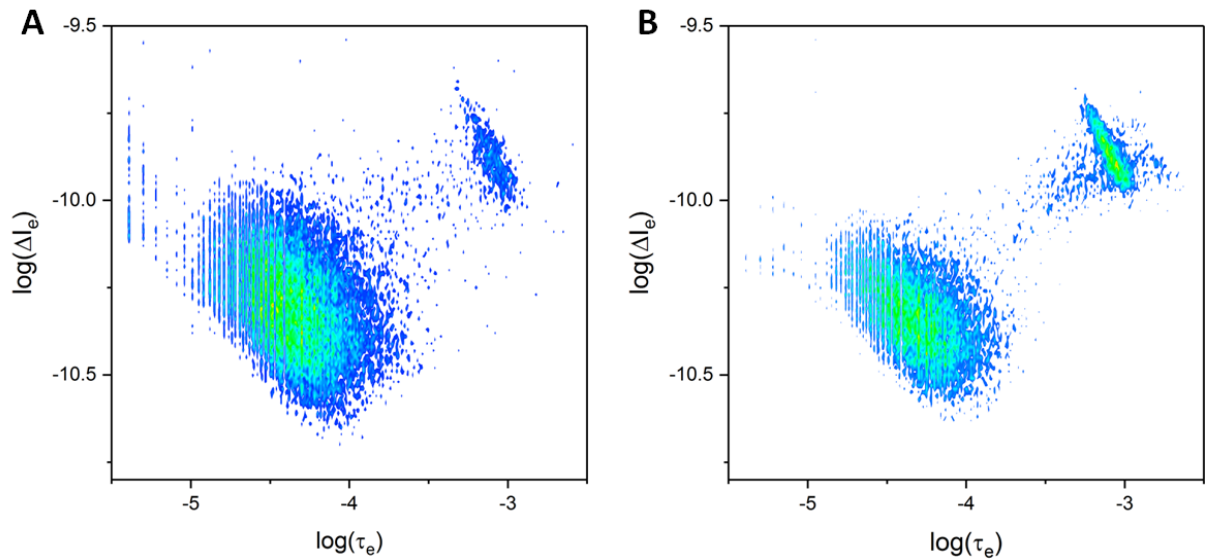

**Fig. S6**: Comparison of translocation data for hyx1 (A) and ssx1 (B). The data in panel A are from three different pipettes (59964 events in total), as described in the main manuscript; those in panel B from four pipettes (29776 events in total). DNA fragments are seen in both cases, in the larger cluster (bottom left), while the DNA carrier forms a smaller, separate cluster (top right). The translocation characteristics for the two samples, and in particular the event cluster, are virtually the same, suggesting that hybridisation of the overhang does not influence the translocation of the carrier significantly.

## 5) Sub-event translocation data for $V_{bias}$ = 0.5 V

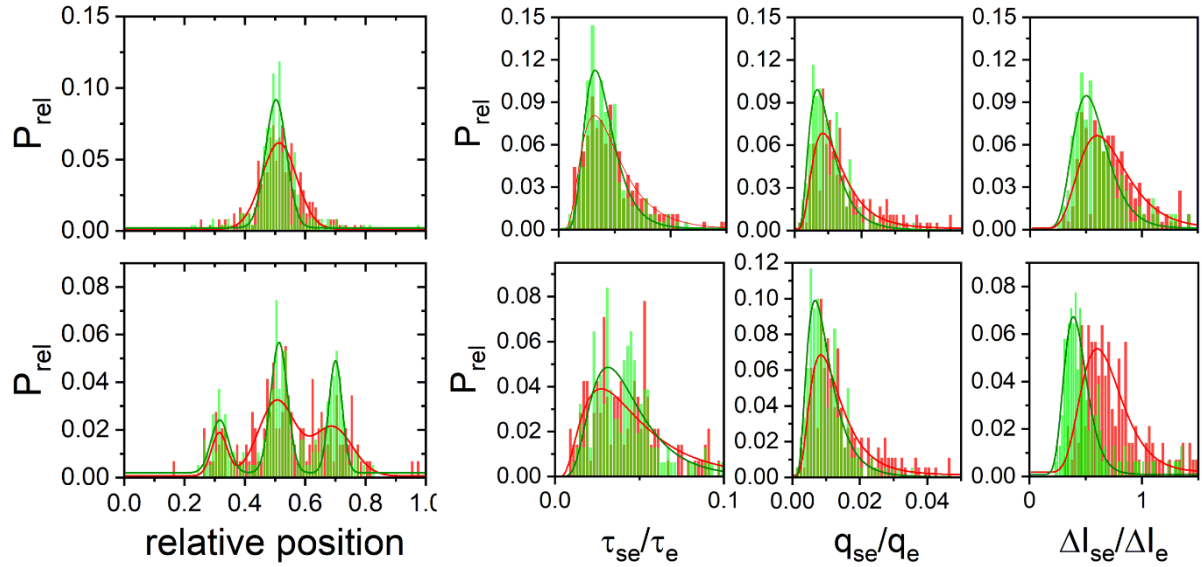

**Fig. S7**: DNA carrier translocation data, analysis of sub-events ($V_{bias}$ = 0.5 V). The arrangement and color-coding is the same as in fig. 4 of the main text, namely ssx1/hyx1 (top) and ssx2/hyx2 (bottom). ssx samples in green, hyx in red. Left: Normalised histograms of the relative sub-event position for single-overhang samples (top) and the double-overhang samples (bottom), inc. Gaussian fits. Right: Normalised histograms of sub-event characteristics, relative to the respective event: $\tau_{se}/\tau_e$, $q_{se}/q_e$ and $\Delta I_{se}/\Delta I_e$. All distributions are non-Gaussian and are represented well by log-normal fits (solid lines). As for $V_{bias}$ = 0.7 V, the largest difference between the ssx and hyx samples appears to be in the $\Delta I_{se}/\Delta I_e$ distributions, which were then used for further analysis. We note that the experiments at different $V_{bias}$ were performed in different pipettes, so the data from the two voltage conditions are independent.

## 6) Results of normality testing (Kolmogorov-Smirnow (K.S.) test)

| Sample | DF | Statistic | p-value | Decision at level (5%) |
|---|---|---|---|---|
| hyx1, 0.7 V | 367 | 0.05602 | 0.19736 | Can't reject normality |
| ssx1, 0.7 V | 496 | 0.07127 | 0.01271 | Reject normality |
| hyx1, 0.5 V | 181 | 0.06954 | 0.34066 | Can't reject normality |
| ssx1, 0.5 V | 180 | 0.05942 | 0.55318 | Can't reject normality |
| hyx2, 0.7 V | 399 | 0.03701 | 0.66657 | Can't reject normality |
| ssx2, 0.7 V | 435 | 0.06116 | 0.07606 | Can't reject normality |
| hyx2, 0.5 V | 141 | 0.08594 | 0.24187 | Can't reject normality |
| ssx2, 0.5 V | 155 | 0.10657 | 0.05654 | Can't reject normality |

**Table TS1**: Results from the K.S. test for the individual samples. 'DF': Degrees of Freedom; 'Statistic': K.S. test statistic. Normality is only formally rejected in one case, ssx1 at $V_{bias}$ = 0.7 V.

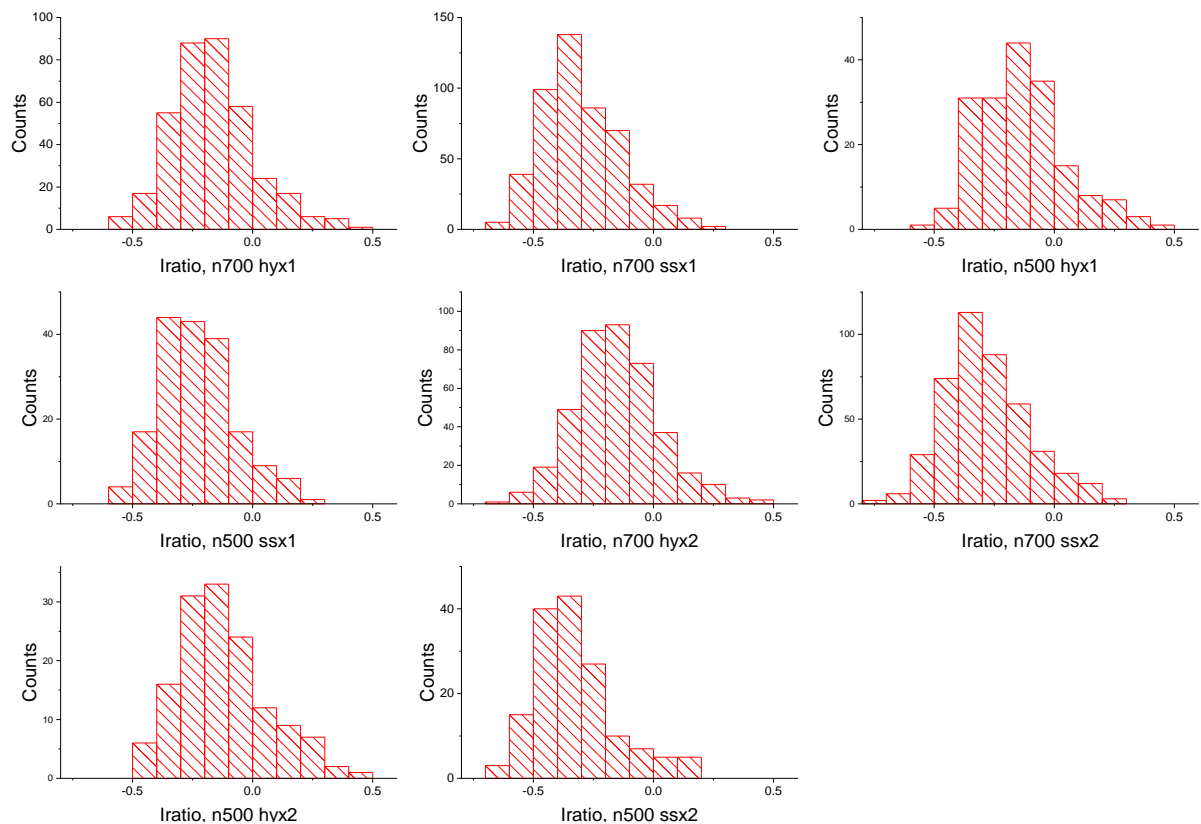

**Fig. S8**: $\log(\Delta I_{se}/\Delta I_e)$ histograms for the individual samples. Despite log-transformation, they still are slightly asymmetric. However, given that ANOVA is known to be rather robust towards non-normality, this does not (significantly) affect the conclusions form the analysis.

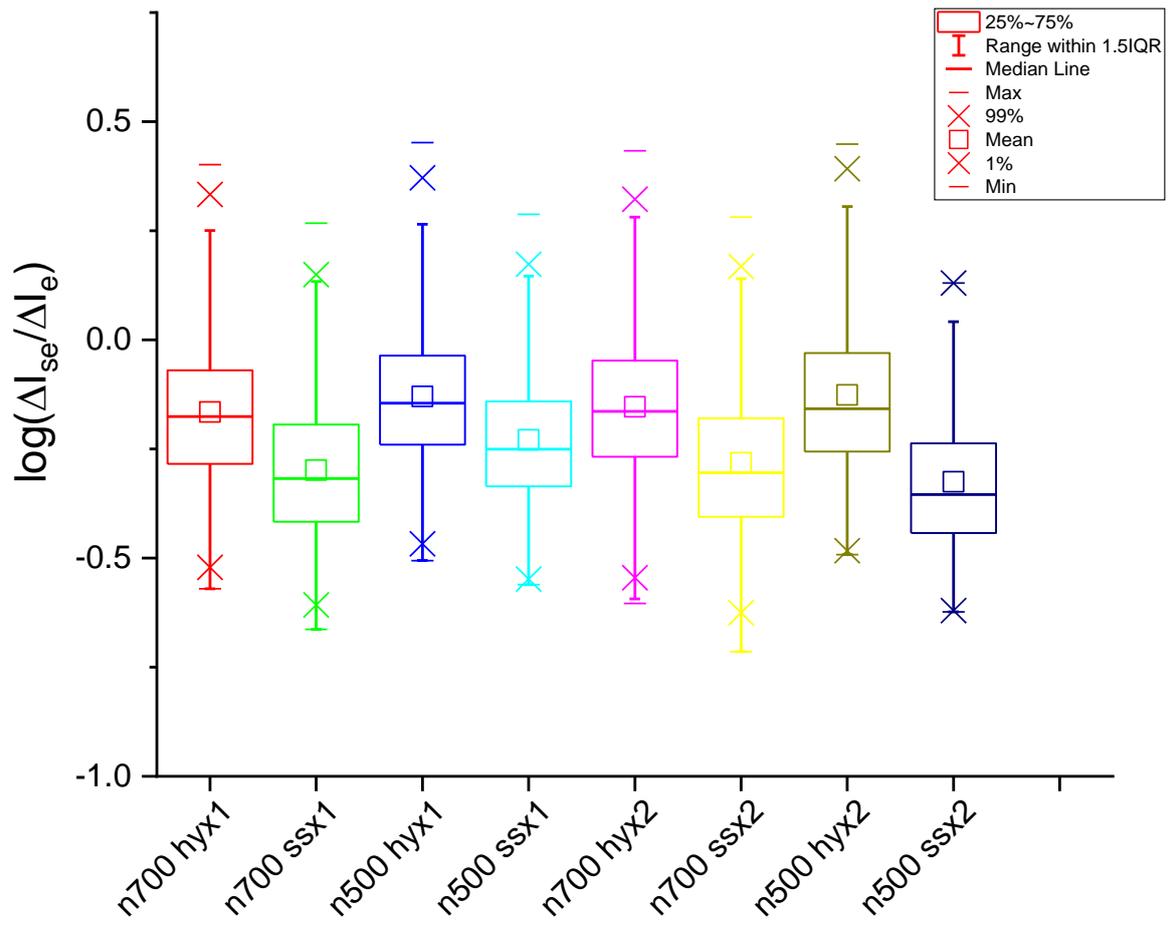

**Fig. S9**: Box plot corresponding to the datasets shown above.

**7) Further discussion of the ANOVA results, main and interaction effects**

The results of the three-factor ANOVA are shown in fig. S10 below, according to the three factors from top to bottom. The black squares represent the mean values of $\log(\Delta I_{es}/\Delta I_e)$, the individual data points for each set are shown as colormaps (white/blue: low/high density). The means of the first two factors, 'hybridisation state' and 'voltage', are significantly different at a 0.05 confidence level (p ≈ 0 and 0.008; means difference: -0.139 and 0.021; sample sizes: 1266/1088 and 657/1697), while the third one, 'number of overhangs', is not (p = 0.061; means difference: 0.015; sample size: 1224/1130). We note that the sample sizes are approximately equal in most cases, except for the effect of 'voltage', which however does not affect the homogeneity of variance test, see main text. Notably, the first observation is of significant interest in that it seems to support the underlying idea of the present work. However, the interpretation of these main effects is only straightforward in the absence of significant interaction terms. Such interaction terms represent the effect of one factor on another one, for example if there were a significant difference between ssx and hyx for one $V_{bias}$ value, but not for another one. In that case, $V_{bias}$ would have to be considered explicitly, in order to make a statement regarding the statistical significance of the difference between ssx and hyx (and so forth). In the presence of significant interaction, suitable *post hoc* tests are needed to explore the relation between different sub-groups, as we show in the main text. The overall ANOVA results are tabulated below.

|  | DF | Sum of Squares | Mean Square | F Value | P Value | Sig |
|---|---|---|---|---|---|---|
| Voltage | 1 | 0.2052 | 0.2052 | 7.080 | 0.0079 | 1 |
| Sample | 1 | 9.0035 | 9.0035 | 310.666 | 0 | 1 |
| #sub-ev. | 1 | 0.1016 | 0.1016 | 3.506 | 0.0613 | 0 |
| Voltage·Sample | 1 | 0.0290 | 0.0290 | 0.9999 | 0.3175 | 0 |
| Voltage·#sub-ev. | 1 | 0.3975 | 0.3975 | 13.714 | 2.177E-4 | 1 |
| Sample·#sub-ev. | 1 | 0.2892 | 0.2892 | 9.9797 | 0.0016 | 1 |
| Voltage·Sample·#sub-ev. | 1 | 0.3459 | 0.3459 | 11.935 | 5.606E-4 | 1 |
| Model | 7 | 11.834 | 1.6906 | 58.335 | 0 |  |
| Error | 2346 | 67.990 | 0.0290 | 0 | 0 |  |
| Corrected Total | 2353 | 79.824 | 0 | 0 | 0 |  |

**Table TS2**: ANOVA results table (overall). Significant interaction are labelled '1' in the last column.

Accordingly, at the 0.05 confidence level, we found two significant 'first-order' interactions, namely between 'voltage' and 'number of overhangs' ('Voltage·#sub-ev.'; F-value: 13.71; P-value: 2.177·10$^{-4}$) and 'hybridisation state' and 'number of overhangs' ('sample·#sub-ev.'; F-value: 9.980; P-value: 0.0016), respectively. In the first case, this would suggest that the statistical significance of the difference in the $\log(\Delta I_{se}/\Delta I_e)$ distributions between single- and double-overhang samples could depend on voltage. From a sensing point of view, it is clearly desirable for there to be no interaction with the number of overhangs, because it implies that then they are detected independently from each other. Equally, in the second case, the result suggests that depending on the number of overhangs, the statistical significance in differentiating ssx from hyx samples may vary.

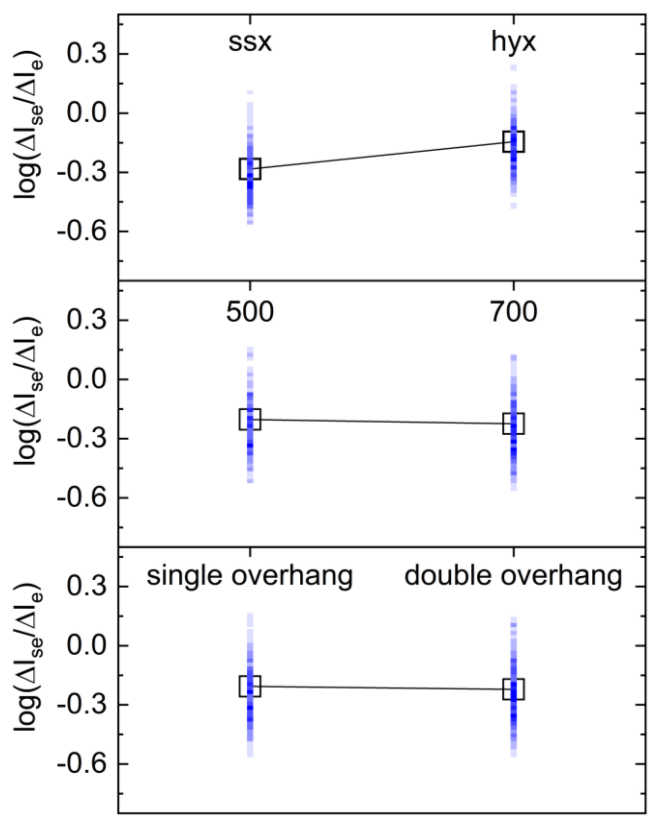

**Fig. S10**: Results for a 3-way ANOVA on the log-transformed $\Delta I_{se}/\Delta I_e$ data. The factors and levels were the hybridisation state ('ssx', 'hyx'; top), the voltage (500 mV, 700 mV; middle) and the number of overhangs/ protrusions on the DNA carrier ('single', 'double'; bottom). Black squares: Mean values. Colormaps: Histograms of the individual $\log(\Delta I_{se}/\Delta I_e)$ values with blue and white corresponding to high and low density, respectively. The means of the first two factors are significantly different at a 0.05 confidence level (p = 0 and 0.008), while the third is not (p = 0.061). See main text for further discussion.

Starting with the interaction between 'voltage' and 'number of overhangs', it is apparent from the table below that the difference between ssx and hyx samples is statistically significant for $V_{bias}$ = 0.5 V (at the 0.05 level; means difference = +0.0438), while at $V_{bias}$ = 0.7 V it is not (means difference = -0.0144). Thus, the reason why the main effect of 'number of overhangs' (single vs. double) is not found to be statistically significant, is most likely due to the partial cancellation of the means differences between different datasets. More detailed investigation further reveals that in fact only the difference between ssx1 and ssx2 at $V_{bias}$ = 0.5 V is statistically significant (means difference: +0.0959), while hyx1 vs. hyx2 at $V_{bias}$ = 0.5 V, and ssx1 vs. ssx2 and hyx1 vs. hyx2 at $V_{bias}$ = 0.7 V are not (means differences: -0.0082; -0.0167; -0.0121, respectively). Therefore, at the individual level there does not seem to be a consistent statistically significant difference between the single and double-overhang samples, in line with the overall main effect.

A slightly different picture emerges for the second observed interaction, namely between 'hybridisation state' and 'number of overhangs'. A subset of the underlying direct comparisons, namely ssx1 vs. hyx1 and ssx2 vs. hyx2, represent the core of the present study and are discussed in detail in the main text. The other comparisons, ssx1 vs. hyx2 and ssx2 vs. hyx1, also yield statistically significant differences, but in absence of a statistically significant effect from the number of overhangs, they simply reflect differences in hybridisation state. In the context of the present study, they are less relevant, as we aim to determine the hybridisation state of a given DNA structure.

| Voltage | Sample | #sub-ev. | Voltage | Sample | #sub-ev. | MeanDiff | SEM | q Value | Prob | Sig |
|---|---|---|---|---|---|---|---|---|---|---|
| 500 | ssx | -- | 500 | hyx | -- | -0.14653 | 0.01333 | -15.541 | 2.22E-16 | 1 |
| 500 | ssx | -- | 700 | ssx | -- | 0.01307 | 0.01086 | 1.702 | 0.62 | 0 |
| 500 | ssx | -- | 700 | hyx | -- | -0.11773 | 0.01116 | -14.924 | 2.22E-16 | 1 |
| 500 | hyx | -- | 700 | ssx | -- | 0.1596 | 0.01106 | 20.413 | 2.22E-16 | 1 |
| 500 | hyx | -- | 700 | hyx | -- | 0.0288 | 0.01135 | 3.588 | 0.05 | 0 |
| 700 | ssx | -- | 700 | hyx | -- | -0.1308 | 0.0083 | -22.280 | 2.22E-16 | 1 |
| 500 | -- | single | 500 | -- | double | 0.04387 | 0.01333 | 4.652 | 0.00554 | 1 |
| 500 | -- | single | 700 | -- | single | 0.05007 | 0.01069 | 6.625 | 1.67E-5 | 1 |
| 500 | -- | single | 700 | -- | double | 0.03567 | 0.01071 | 4.709 | 0.0048 | 1 |
| 500 | -- | double | 700 | -- | single | 0.0062 | 0.01149 | 0.763 | 0.9493 | 0 |
| 500 | -- | double | 700 | -- | double | -0.0082 | 0.01151 | -1.008 | 0.8922 | 0 |
| 700 | -- | single | 700 | -- | double | -0.0144 | 0.0083 | -2.453 | 0.3055 | 0 |
| -- | ssx | single | -- | ssx | double | 0.03958 | 0.01086 | 5.157 | 0.0015 | 1 |
| -- | ssx | single | -- | hyx | single | -0.11381 | 0.01069 | -15.0590 | 2.220E-16 | 1 |
| -- | ssx | single | -- | hyx | double | -0.12393 | 0.01113 | -15.741 | 2.220E-16 | 1 |
| -- | ssx | double | -- | hyx | single | -0.1534 | 0.01108 | -19.580 | 2.220E-16 | 1 |
| -- | ssx | double | -- | hyx | double | -0.16352 | 0.01151 | -20.090 | 2.220E-16 | 1 |
| -- | hyx | single | -- | hyx | double | -0.01012 | 0.01135 | -1.261 | 0.809 | 0 |
| 500 | ssx | single | 500 | ssx | double | 0.0959 | 0.01862 | 7.283 | 7.218E-6 | 1 |
| 500 | ssx | single | 500 | hyx | single | -0.0945 | 0.01789 | -7.471 | 3.534E-6 | 1 |
| 500 | ssx | single | 500 | hyx | double | -0.10267 | 0.01911 | -7.597 | 2.167E-6 | 1 |
| 500 | ssx | single | 700 | ssx | single | 0.06938 | 0.01479 | 6.635 | 7.377E-5 | 1 |
| 500 | ssx | single | 700 | ssx | double | 0.05265 | 0.01506 | 4.944 | 0.011 | 1 |
| 500 | ssx | single | 700 | hyx | single | -0.06374 | 0.01546 | -5.829 | 9.807E-4 | 1 |
| 500 | ssx | single | 700 | hyx | double | -0.07582 | 0.01526 | -7.027 | 1.853E-5 | 1 |
| 500 | ssx | double | 500 | hyx | single | -0.1904 | 0.0186 | -14.478 | 4.441E-16 | 1 |
| 500 | ssx | double | 500 | hyx | double | -0.19856 | 0.01978 | -14.198 | 4.441E-16 | 1 |
| 500 | ssx | double | 700 | ssx | single | -0.02652 | 0.01564 | -2.398 | 0.690 | 0 |
| 500 | ssx | double | 700 | ssx | double | -0.04325 | 0.0159 | -3.847 | 0.116 | 0 |
| 500 | ssx | double | 700 | hyx | single | -0.15964 | 0.01628 | -13.868 | 4.441E-16 | 1 |
| 500 | ssx | double | 700 | hyx | double | -0.17172 | 0.01609 | -15.098 | 4.441E-16 | 1 |
| 500 | hyx | single | 500 | hyx | double | -0.00817 | 0.01909 | -0.605 | 0.9999 | 0 |
| 500 | hyx | single | 700 | ssx | single | 0.16388 | 0.01476 | 15.704 | 4.441E-16 | 1 |
| 500 | hyx | single | 700 | ssx | double | 0.14715 | 0.01503 | 13.844 | 4.441E-16 | 1 |
| 500 | hyx | single | 700 | hyx | single | 0.03076 | 0.01544 | 2.818 | 0.487 | 0 |
| 500 | hyx | single | 700 | hyx | double | 0.01868 | 0.01523 | 1.734 | 0.924 | 0 |
| 500 | hyx | double | 700 | ssx | single | 0.17205 | 0.01622 | 15.001 | 4.441E-16 | 1 |
| 500 | hyx | double | 700 | ssx | double | 0.15532 | 0.01647 | 13.337 | 4.441E-16 | 1 |
| 500 | hyx | double | 700 | hyx | single | 0.03892 | 0.01684 | 3.269 | 0.287 | 0 |
| 500 | hyx | double | 700 | hyx | double | 0.02684 | 0.01665 | 2.280 | 0.743 | 0 |
| 700 | ssx | single | 700 | ssx | double | -0.01673 | 0.01116 | -2.119 | 0.809 | 0 |
| 700 | ssx | single | 700 | hyx | single | -0.13313 | 0.0117 | -16.089 | 4.441E-16 | 1 |
| 700 | ssx | single | 700 | hyx | double | -0.1452 | 0.01143 | -17.967 | 4.441E-16 | 1 |
| 700 | ssx | double | 700 | hyx | single | -0.1164 | 0.01205 | -13.665 | 4.441E-16 | 1 |
| 700 | ssx | double | 700 | hyx | double | -0.12847 | 0.01178 | -15.423 | 4.441E-16 | 1 |
| 700 | hyx | single | 700 | hyx | double | -0.01208 | 0.01229 | -1.390 | 0.977 | 0 |

**Table TS3**: ANOVA results table of interaction effects ($\alpha$ = 0.05 in all cases).

## 8) Representation of the translocation data shown in fig. 3 on linear scales

For comparison, we show the data displayed in fig. 3 of the main manuscript, but now on linear scales. Short translocation events are markedly bunched on the left-hand side, which illustrates the advantage of using a logarithmic representation.

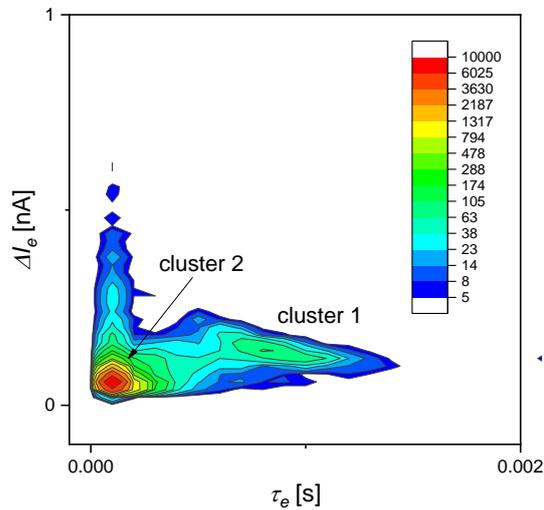

**Fig. S11:** Translocation data from three different nanopipettes for hyx1 ($V_{bias}$ = 0.7 V, 4 M LiCl + 10 mM TE electrolyte), as in fig. 3.

## 9) AFM imaging of the hyx1 construct, further examples

Below we show some further examples of AFM images of hyx1, cf. fig. 2 in the main manuscript. In some cases, the overhang is easily identified, by visual inspection. In other cases, height information had to be taken into account as well, for example to rule out artefacts from DNA knots or coiling, as described in the Methods section of the main manuscript.

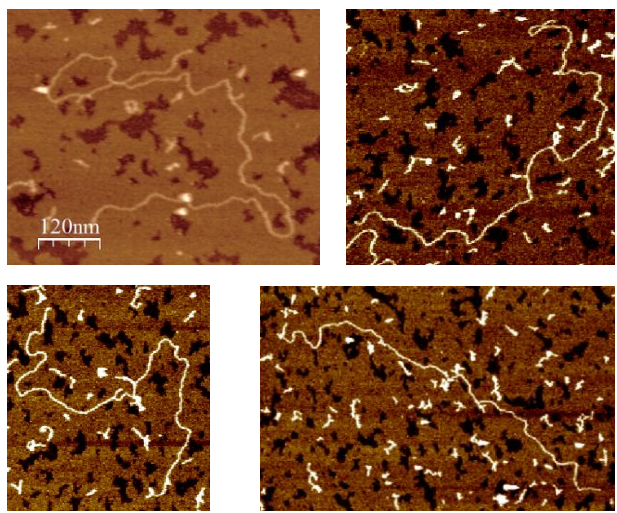

**Fig. S12**: AFM imaging of the hyx1 construct, further examples.

| DNA sample | First segment length [nm] | Overhang length [nm] | Second segment length [nm] | Total length [nm] (wo. overhang) |
|---|---|---|---|---|
| 1 | 979 | 29.6 | 903 | 1882 |
| 2 | 903 | 50.8 | 1064 | 1967 |
| 3 | 922 | 38.8 | 981 | 1903 |
| 4 | 895 | 34.8 | 990 | 1885 |
| 5 | 760 | 45 | 778 | 1538 |
| 6 | 859 | 34.4 | 906 | 1765 |
| 7 | 970 | 39.2 | 972 | 1942 |
| 8 | 955 | 31 | 899 | 1854 |
| 9 | 853 | 25.9 | 830 | 1683 |
| 10 | 966 | 36.4 | 765 | 1731 |
| 11 | 725 | 42.1 | 951 | 1676 |
| 12 | 1090 | 47 | 849 | 1939 |
| 13 | 971 | 44.1 | 952 | 1923 |
| 14 | 839 | 46.8 | 1010 | 1849 |
| 15 | 818 | 24.8 | 796 | 1614 |
| 16 | 1210 | 48 | 717 | 1927 |
| 17 | 881 | 45 | 887 | 1768 |
| 18 | 766 | 31 | 983 | 1749 |
| 19 | 1116 | 42.7 | 703 | 1819 |
| 20 | 1020 | 32 | 897 | 1917 |
| 21 | 955 | 71.7 | 934 | 1889 |
| 22 | 979 | 36.6 | 673 | 1652 |
| 23 | 816.5 | 40.5 | 777 | 1593.5 |
| | | | | |
| Average: | | 39.2 | | 1849 |

**Table TS4**: Length data forming the basis of the histograms shown in fig. 2C in the main manuscript. Sample 1 is shown in fig. 2B, samples 14-17 are shown in fig. S12 above.